\documentclass[12pt]{article}
\usepackage{amsfonts}

\begin{document}

\title{Gauge strata and particle generations}
\author{R.Vilela Mendes \\
{\small Grupo de F\'{\i }sica Matem\'{a}tica}\\
{\small \ Complexo Interdisciplinar, Universidade de Lisboa,}\\
{\small \ Av. Gama Pinto, 2 - P1699 Lisboa Codex, Portugal}}
\date{}
\maketitle

\begin{abstract}
Phenomenological evidence suggests the existence of non-trivial background
fields in the QCD vacuum. On the other hand SU(3) gauge theory possessses
three different classes of both non-generic and non-trivial strata that may
be used as classical backgrounds. It is suggested that this three-fold
multiplicity of non-trivial vacua may be related to the existence of
particle generations, which would then find an explanation in the framework
of the standard model.
\end{abstract}


Within the limits of its accuracy, all experimental data known so far is
consistent with a $SU(3)_{c}\times SU(2)_{L}\times U(1)_{Y}$ gauge theory as
a model for particle physics. However, several features of the observed low
energy phenomena remain to be explained. Among these the generation
structure, in the quark and lepton spectra, stands as one of the most
intriguing puzzles.

Several schemes, going beyond the gauge group of the standard model, have
been proposed to describe the existence of particle generations, ranging
from family groups, horizontal symmetries, radial excitations, enlarged
groups, preon models to superstrings.

Contrary to this \textit{beyond the standard model }trend, I will argue in
this letter that, by itself, the gauge group of the standard model already
contains a multiplicity in the space of its solutions that is reminiscent of
the particle generations structure.

For definiteness I will consider physical states as being represented by
quantum fluctuations around classical solutions and physical processes as
path integrals on the space of field configurations. A classical gauge
theory consists of four basic objects:

(i) A principal fiber bundle $P\left( M,G\right) $ with structure group $G$
and projection $\pi :P\rightarrow M$,

(ii) An affine space $\mathcal{C}$ of connections on $P$ which, by selecting
a reference connection, may be modelled by a vector space $\mathcal{A}$ of
one-forms on $M$ with values on the Lie algebra $\mathcal{G}$ of $G$,

(iii) The space of differentiable sections of $P$, called the \textit{gauge
group }$\Gamma $

(iv) A $\Gamma -$invariant functional $\mathcal{L}:\mathcal{A}\rightarrow 
\mathbb{R}$

All statements below refer to the case where $G$ is a compact group. The
action of $\Gamma $ on $\mathcal{A}$ leads to a stratification of $\mathcal{A%
}$ corresponding to the classes of equivalent \textit{orbits} $\left\{
A^{g};g\in \Gamma \right\} $. Let $\Gamma _{A}$ denote the \textit{isotropy
group} of $A\in \mathcal{A}$%
\begin{equation}
\Gamma _{A}=\left\{ g\in \Gamma :A^{g}=A\right\}   \label{1}
\end{equation}
The \textit{stratum} $\Sigma \left( A\right) $ of $A$ is the set of
connections having isotropy groups $\Gamma -$conjugated to that of $A$%
\begin{equation}
\Sigma \left( A\right) =\left\{ B\in \mathcal{A}:\exists g\in \Gamma :\Gamma
_{B}=g\Gamma _{A}g^{-1}\right\}   \label{2}
\end{equation}
The configuration space of the gauge theory is the quotient space $\mathcal{A%
}/\Gamma $ and therefore a stratum is the set of points in $\mathcal{A}%
/\Gamma $ that correspond to orbits with conjugated isotropy groups.

The stratification of the gauge space when $G$ is a compact group has been
extensively studied\cite{Kondracki1} - \cite{Rudolph}. The stratification is
topologically regular. The set of strata carries a partial ordering of
types, $\Sigma _{\tau }\subseteq \Sigma _{\tau ^{^{\prime }}}$ with $\tau
\leq \tau ^{\prime }$ if there are representatives $S_{\tau }$ and $S_{\tau
^{\prime }}$ of the isotropy groups such that $S_{\tau }\supseteq S_{\tau
^{\prime }}$. The maximal element in the ordering of types is the class of
the center $Z(G)$ of $G$ and the minimal one is the class of $G$ itself.
Furthermore $\cup _{t\geq \tau }\Sigma _{t}$ is open and $\Sigma _{\tau }$
is open in the relative topology in $\cup _{t\leq \tau }\Sigma _{t}$.

Most of the stratification results have been obtained in the framework of
Sobolev connections and Hilbert Lie groups. However, for the calculation of
physical quantities in the path integral formulation 
\begin{equation}
\left\langle \phi \right\rangle =\int_{\mathcal{A}/\Gamma }\phi \left( \xi
\right) e^{i\mathcal{L}\left( \xi \right) }d\mu \left( \xi \right) 
\label{3}
\end{equation}
a measure in $\mathcal{A}/\Gamma $ is required, and no such measure has been
found for Sobolev connections. Therefore it is more convenient to work in a
space of generalized connections $\overline{\mathcal{A}}$, defining parallel
transports on piecewise smooth paths as simple homomorphisms from the paths
to the group $G$, without a smoothness assumption\cite{Ashtekar1}. The same
applies to the generalized gauge group $\overline{\Gamma }$. Then, there is
in $\overline{\mathcal{A}}/\overline{\Gamma }$ an induced Haar measure, the
Ashtekar-Lewandowski measure\cite{Ashtekar2} - \cite{Ashtekar3}, Sobolev
connections being a dense zero measure subset of the generalized connections%
\cite{Marolf}. The question remained however of whether the stratification
results derived in the context of Sobolev connections would apply to
generalized connections. This question was recently settled by Fleischhack%
\cite{Fleischhack} who, by establishing a slice theorem for generalized
connections, proved that essentially all existing stratification results
carry over to the generalized connections. In some cases they even have
wider generality.

Because the isotropy group of a connection is isomorphic to the centralizer
of its holonomy group\cite{Booss}, the strata are in one-to-one
correspondence with the Howe subgroups of $G$, that is, the subgroups that
are centralizers of some subset in $G$. Given an holonomy group $H_{\tau }$
associated to a connection $A$ of type $\tau $, the stratum of $A$ is
classified by the conjugacy class of the isotropy group $S_{\tau }$, the
centralizer of $H_{\tau }$ 
\begin{equation}
S_{\tau }=Z\left( H_{\tau }\right)   \label{4}
\end{equation}
An important role is also played by the centralizer of the centralizer 
\begin{equation}
H_{\tau }^{\prime }=Z\left( Z\left( H_{\tau }\right) \right)   \label{5}
\end{equation}
that contains $H_{\tau }$ itself. If $H_{\tau }^{\prime }$ is a proper
subgroup of $G$ the connection $A$ reduces locally to the subbundle $P_{\tau
}=\left( M,H_{\tau }^{\prime }\right) $. Global reduction depends on the
topology of $M$, but it is always possible if $P$ is a trivial bundle. $%
H_{\tau }^{\prime }$ is the structure group of the \textit{maximal subbundle}
associated to type $\tau $.Therefore the types of strata are also in
correspondence with types of reductions of the connections to subbundles. If 
$S_{\tau }$ is the center of $G$ the connection is called \textit{irreducible%
}, all others are called \textit{reducible}. The stratum of the irreducible
connections is called the \textit{generic stratum}. It is open and dense and
it carries the full Ashtekar-Lewandowski measure.

Now I turn to the case $G=SU(3)$. The isotropy groups and the structure
groups of the maximal subbundles are\cite{Heil} : 
\begin{equation}
\begin{array}{ccc}
& \Gamma _{A} & H_{A}^{^{\prime }} \\ 
1 & \mathbb{Z}_{3} & SU(3) \\ 
2 & U(1) & U(2) \\ 
3 & U(1)\times U(1) & U(1)\times U(1) \\ 
4 & U(2) & U(1) \\ 
5 & SU(3) & \mathbb{Z}_{3}
\end{array}
\label{6}
\end{equation}
There are five strata. Stratum 1 is the generic stratum. All others are
reducible strata. Recall now the basic assumption that physical states are
represented by quantum fluctuations around classical solutions. Because of
its full measure, quantum fluctuations in the path integral must be taken
from the generic stratum 1. However classical solutions, around which the
quantum fluctuations take place, are not required to belong to the generic
stratum. For example the perturbative vacuum is in the stratum 5, that is
the stratum to which the null connections $\left( A_{\mu }(x)=0\right) $
belong.

One-loop calculations show that the perturbative vacuum is unstable and,
even more important, there is ample phenomenological evidence for the
existence of non-trivial vacuum condensates in the QCD vacuum\cite{Shifman1} 
\cite{Shifman2}. This is incompatible with stratum 5 being the site for the
physical vacuum. Therefore classical vacuum solutions should be looked for
in the other reducible strata.

A classical solution of the gauge theory is a stationary point of the $%
\Gamma -$invariant functional $\mathcal{L}$, that is, a $\mathcal{L-}$%
critical point in $\overline{\mathcal{A}}/\overline{\Gamma }$. Gaeta and
Morando\cite{Gaeta1}, generalizing the classical (finite-dimensional) result
of Michel\cite{Michel1}, have proven that an orbit in $\overline{\mathcal{A}}%
/\overline{\Gamma }$ is critical for any $\Gamma -$invariant functional
whatsoever if and only if it is isolated in its stratum. This applies of
course to the most singular stratum (1), not to the other reducible strata.
However using compactness arguments it is possible to prove the existence of
stationary points for each particular $\Gamma -$invariant functional.
Alternatively, by constructing representatives of the connections in each
one of the strata 2 - 4, it is easy to check that they contain a large
number of stationary points of the Yang-Mills action. Which one is the
minimum energy classical solution is not important for our discussion.
Consistency of the choice of a classical solution in a reducible stratum is
insured by the fact that trajectories of the classical field theory remain
in that same stratum\cite{Otto}.

In conclusion: to be compatible with non-trivial vacuum condensates in QCD,
classical solutions should not be chosen in the minimal stratum (1) and we
are left with a three-fold degeneracy of vacuum possibilities.

Quarks are triplet excitations over these vacua and leptons singlet
excitations. Therefore a three family structure seems to emerge already at
the level of the solutions of the standard model gauge group, without the
need to go beyond this group.

Further insight is obtained when the gauge theory is considered as an
infinite-dimensional symplectic structure, with space components of the
connection and the chromoelectric field as coordinates and momenta, defined
on a Cauchy surface of initial data ($x^{0}=0$ for example)\cite{Arms1} \cite
{Arms2}. The Yang-Mills equations are split into evolution equations and
constraints on the Cauchy surface. The solutions to the whole system form a
fibre bundle over the solutions to the constraint equations. Therefore to
describe the singularities of the full equations it suffices to describe
those of the constraints.

The covariant derivative $\Gamma =D\overrightarrow{E}$ of the chromoelectric
field on the Cauchy surface being the generator of the gauge
transformations, $\Gamma $ defines a momentum mapping and the set $C$ of
solutions of the constraint equations ($\Gamma =0$) is the zero set $\Gamma
^{-1}(0)$ of the momentum mapping.

In the neighborhood of a field (\underline{$\overrightarrow{A}$},\underline{$%
\overrightarrow{E}$}) without symmetries the solution set $C$ is a manifold
with tangent space given by $\ker d\Gamma $, that is 
\begin{equation}
\overrightarrow{\nabla }\cdot \overrightarrow{e}+\left[ \underline{%
\overrightarrow{A}}\overrightarrow{,e}\right] +\left[ \overrightarrow{a},%
\underline{\overrightarrow{E}}\right] =0  \label{7}
\end{equation}
with $\overrightarrow{A}=\underline{\overrightarrow{A}}+\overrightarrow{a}$
and $\overrightarrow{E}=\underline{\overrightarrow{E}}+\overrightarrow{e}$.
This is the case for fields in the generic stratum (stratum 1).

In the neighborhood of fields in all the other strata, the zero set of the
momentum mapping has singularities and the solution set $C$ is diffeomorphic
to the product of a manifold and the zero set of a homogeneous quadratic
function. It means that in addition to the linear constraint, Eq.(\ref{7}),
and a slice condition\cite{Arms1}, fields of the same strata in the
neighborhood of (\underline{$\overrightarrow{A}$},\underline{$%
\overrightarrow{E}$}) must also satisfy a quadratic constraint, which in
components is 
\begin{equation}
f_{bcd}a_{c}^{k}e_{d}^{k}=0  \label{8}
\end{equation}
($f_{bcd}$ being the structure constants of $G$). This is a condition to be
taken into account when quantizing a gauge system around fields of the
non-generic strata. This additional constraint suppresses transitions to
configurations of lower symmetry and, by analogy with the solutions of the
Schr\"{o}dinger equation on double cones\cite{Emmrich}, one expects low
energy states to remain concentrated near the singularities. That is, not
only the classical solutions remain in the same strata, but also low energy
quantum solutions are expected to approximate and inherit the symmetries of
the singular strata.

Of course, all these considerations only point out that, if the trivial
vacuum is unstable, then there is a three-fold choice for non-trivial
reducible backgrounds. Nothing is said about why triplet or singlet
excitations over each type of vacuum have different masses. Here there are
two possibilities. Part of the mass splitting may arise from the diverse
nature of the vacuum classes. Remember that important algebraic and coset
volume differences exist among the reducible connections. In particular the
closed set nature of $\cup _{t\leq \tau }\Sigma _{t}$, implies that each $%
\Sigma _{t}$ behaves like a boundary set for the strata of type immediately
higher. Alternatively, and for the mass contributions that are not sensitive
to the background, a democratic mass matrix with all elements equal might be
considered, as discussed by a number of authors\cite{Fritzsch} \cite{Georgi} 
\cite{Ramond}. When diagonalized this leads (in leading order) to one
massive state and two massless ones.

In conclusion: Whatever dynamical mechanism provides a full explanation of
the mass differences between the particle generations, the fact is that the
rich strata structure of the standard model must be taken into account
whenever the need be felt to consider non-trivial backgrounds.

\end{document}